\documentclass[twocolumn,english,aps,reprint,groupedaddress,notitlepage,nobibnote,nofootinbib,preprintnumbers,showpacs]{revtex4-1}
\pdfoutput=1
\usepackage{lmodern}

\usepackage[T1]{fontenc}
\usepackage[latin9]{inputenc}
\usepackage{geometry}
\usepackage{color}
\usepackage{babel}
\usepackage{amsmath}
\usepackage{amssymb}
\usepackage{graphicx}
\usepackage{esint}
\usepackage{wasysym}
\usepackage{array}
\usepackage{comment}
\usepackage{tabu}
\usepackage{relsize}

\usepackage[table]{xcolor}
\usepackage[unicode=true,pdfusetitle,
 bookmarks=true,bookmarksnumbered=false,bookmarksopen=false,
 breaklinks=false,pdfborder={0 0 1},backref=false,colorlinks=true]
 {hyperref}

\hypersetup{
 citecolor=blue,linkcolor=blue,urlcolor=blue}
\makeatletter

 \@ifundefined{textcolor}{}
 {
   \definecolor{BLACK}{gray}{0}
   \definecolor{WHITE}{gray}{1}
   \definecolor{RED}{rgb}{1,0,0}
   \definecolor{GREEN}{rgb}{0,1,0}
   \definecolor{BLUE}{rgb}{0,0,1}
   \definecolor{CYAN}{cmyk}{1,0,0,0}
   \definecolor{MAGENTA}{cmyk}{0,1,0,0}
   \definecolor{YELLOW}{cmyk}{0,0,1,0}
 }
\usepackage{babel}
\newcommand{\be}{\begin{equation}}
\newcommand{\ee}{\end{equation}}
\newcommand{\bea}{\begin{eqnarray}}
\newcommand{\eea}{\end{eqnarray}}

\newcommand{\Fig}[1]{Fig.~\ref{#1}}
\newcommand{\Eq}[1]{Eq.~(\ref{#1})}

\newcommand{\Sec}[1]{Sec.~\ref{#1}}

\definecolor{zero1}{rgb}{0.35,0.4,0.85}
\definecolor{zero2}{rgb}{0.88,0.88,.88}
\definecolor{zero3}{rgb}{0.85,0.85,0.95}
\definecolor{zero4}{rgb}{1,0.3,0.3}

\newcommand{\eq}[1]{\begin{align}#1\end{align}}

\newcommand{\nn}{\nonumber}

\makeatletter
\def\CT@@do@color{%
	\global\let\CT@do@color\relax
		\@tempdima\wd\z@
		\advance\@tempdima\@tempdimb
		\advance\@tempdima\@tempdimc
		\advance\@tempdimb\tabcolsep
		\advance\@tempdimc\tabcolsep
		\advance\@tempdima1.5\tabcolsep
	\kern-1.5\@tempdimb
	\leaders\vrule
	\hskip\@tempdima\@plus  1fill
	\kern-1.5\@tempdimc
	\hskip-\wd\z@ \@plus -1fill }
\makeatother

\begin{document}

%\preprint{\hbox{CALT-2015-024} }

\title{Broad Diphotons from Narrow States}

\author{Haipeng An}
\author{Clifford Cheung}
\author{Yue Zhang}

\affiliation{Walter Burke Institute for Theoretical Physics \\California Institute of Technology, Pasadena, CA 91125\,}
\date{\today}
\email{anhp@caltech.edu, \\ clifford.cheung@caltech.edu, \\ yuezhang@theory.caltech.edu}
%\pacs{}
\begin{abstract}
ATLAS and CMS have each reported a modest diphoton excess consistent with the decay of a broad resonance at $\sim 750$ GeV.  We show how this signal can arise in a weakly coupled theory comprised solely of narrow width particles.  In particular, if the decaying particle is produced off-shell, then the associated diphoton resonance will have a broad, adjustable width.  We present simplified models which explain the diphoton excess through the three-body decay of a scalar or fermion.
Our minimal ultraviolet completion is a weakly coupled and renormalizable theory of a singlet scalar plus a heavy vector-like quark and lepton.
The smoking gun of this mechanism is an asymmetric diphoton peak recoiling against missing transverse energy, jets, or leptons.
\end{abstract}

\preprint{CALT-TH-2015-065}

\maketitle

\section{Introduction\label{intro}}

The ATLAS and CMS collaborations have both observed a curious excess in the diphoton invariant mass spectrum in the vicinity of $\sim $ 750 GeV~\cite{ATLAS, CMS}.  The significance of the excess in each experiment is $\sim$ 3.6$\sigma$ and $\sim$ 2.6$\sigma$, respectively, with a rate consistent with a signal cross-section of $\sim$ 5--10 fb.  Current observations favor a broad resonance, with ATLAS reporting a width of $\sim$ 45 GeV, corresponding to $6\%$ of the mass.

While the diphoton excess may well be a statistical fluke or experimental systematic, it is nevertheless important to understand the range of new physics scenarios which can accommodate this observation. Recently, there has been an enormous flurry of papers discussing the implications for beyond the standard model physics~\cite{theychaseanomaly}, mostly observing that the excess is consistent with a strongly coupled sector linking a new decaying state to gluons and photons with quite large couplings.  

%While this appears to disfavor weakly coupled dynamics, we would like to point out an broad class of exceptions to this intuition.

In this paper we point out that weakly coupled physics can explain both the rate of the diphoton excess as well as its apparently broad width.   To do so, we introduce a now familiar ingredient: a singlet scalar which decays to diphoton.  Such a state is narrow in a weakly coupled theory but if it is predominantly produced {\it off-shell} then the associated diphoton spectrum will be broadened.  This occurs, for example, if the scalar is produced through the three-body decay of a lighter state.   The observed width of the diphoton resonance is then an adjustable parameter controlled by the spectrum of masses.  

A generic prediction of this mechanism is that the peak in the diphoton invariant mass is asymmetric, with more events below the apex than above.  Moreover, the high energy photons in each event should be associated with recoiling missing transverse energy, jets, or leptons.

In \Sec{sec:setup} we summarize the basic mechanism and then analyze a set of simplified models characterizing the three-body decay of a scalar or fermion.  We then present explicit models for the diphoton excess in \Sec{sec:DM} and \Sec{sec:QL}.  Our minimal ultraviolet complete theory is a renormalizable model of a singlet scalar plus a vector-like quark and lepton.  We conclude in \Sec{sec:conclusions}.

\begin{figure}[t]
\vspace*{0.3cm}
\begin{center}
\includegraphics[width=.3\textwidth]{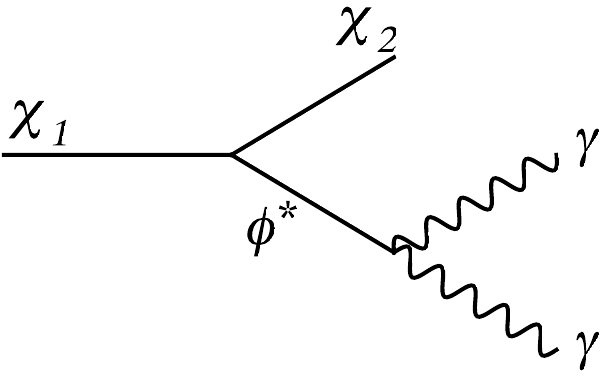}
\end{center}
\vspace*{-0.3cm}
\caption{Three-body decay yielding a slightly off-shell resonance with an effectively adjustable width.}
\label{fig:3bd} 
\end{figure}

\begin{figure*}[t]
\begin{center}
\includegraphics[width=.9\textwidth]{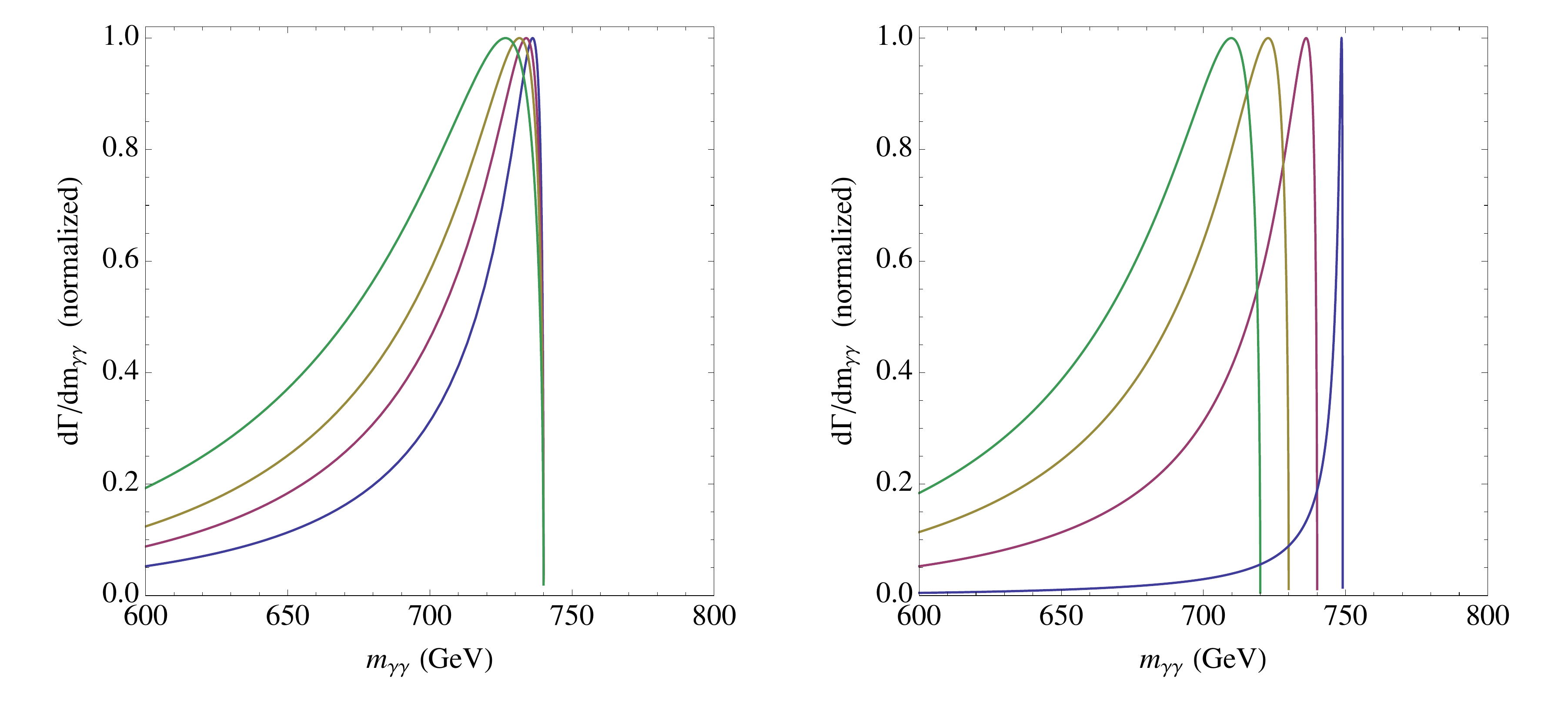}
\end{center}
\vspace*{-0.5cm}
\caption{Differential decay rate for the scalar three-body decay $\chi_1 \rightarrow \chi_2 (\phi^* \rightarrow \gamma\gamma)$  as a function of the diphoton invariant mass, $m_{\gamma\gamma} \equiv |p|$, with maximum normalized to 1. 
In the left panel, $m_{\chi_1}=750$ GeV, $m_{\chi_2}=10$ GeV, and 
$m_\phi-m_{\chi_1}=0$ (blue), 5 (magenta), 10 (yellow), 20 (green) GeV .
In the right panel, $m_{\chi_1}=m_\phi=750\,$GeV, $m_{\chi_2}=1$ (blue), 10 (magenta), 20 (yellow), 30 (green) GeV.
The width of the peak can be adjusted with the particles masses involved in the three-body decay. }
\label{fig:sc} 
\end{figure*}

\section{Basic Setup}

\label{sec:setup}

Our basic setup involves three states: $\phi$, $\chi_1$, and $\chi_2$.  Here $\phi$ is a scalar that decays via $
\phi \rightarrow \gamma \gamma$, although $\phi\rightarrow gg$ may be allowed as well. 
Meanwhile,  $\chi_1$ is created in proton-proton collisions, either directly or from cascade decays.  Once produced, $\chi_1$ decays three-body via
\eq{
\chi_1 \rightarrow \chi_2 (\phi^* \rightarrow \gamma \gamma) \ ,
\label{3body}
}
through an off-shell $\phi$, as shown in \Fig{fig:3bd}.  

For \Eq{3body} to be the dominant decay mode requires
%\begin{itemize}
\eq{
m_\phi &> m_{\chi_1} - m_{\chi_2} \gg 0 \ .
\label{eq:primarycondition}
}   
The first inequality is necessary to forbid the direct two-body decay, $\chi_1 \rightarrow \chi_2 \phi$.  The second inequality is required so the three-body decay $\chi_1 \rightarrow \chi_2 (\phi^* \rightarrow \gamma \gamma)$ into highly energetic photons is kinematically allowed.

Note also that the width of $\phi$ must be of order or greater than the width of $\chi_2$, so the dominant three-body decay is via off-shell $\phi$ rather than off-shell $\chi_2$.   This is automatic if $\chi_2$ is a stable or long-lived particle.  As we will see, this is naturally accommodated if $\chi_1$ and $\chi_2$ are odd under a stabilizing parity.

In the on-shell limit, $m_\phi = m_{\chi_1} - m_{\chi_2}$, the $\phi$ particle is produced at rest, yielding a narrow diphoton resonance regulated by a narrow Breit-Wigner width.  However, as one detunes to the off-shell regime $m_\phi > m_{\chi_1} - m_{\chi_2}$, this pushes the $\phi$ propagator off-shell, widening the peak.

%\sesection{Simplified Models}

%We now present simplified models characterizing the three-body decay required for a broad diphoton spectrum.  These modular constructions can be attached to the bottom of any cascade of new physics states.

%We have remained agnostic about the identity of $\phi_1$, $\phi_2$, and $\phi_3$, but later we will see that some may even be standard model states. 

 In the subsequent sections we study this mechanism with the aid simplified models.  We begin with a discussion of the couplings of $\phi$ to photons and gluons, followed by an analysis of scalar and fermion simplified models for the interactions of $\phi$, $\chi_1$, and $\chi_2$.

\begin{figure*}[t]
\begin{center}
\includegraphics[width=.9\textwidth]{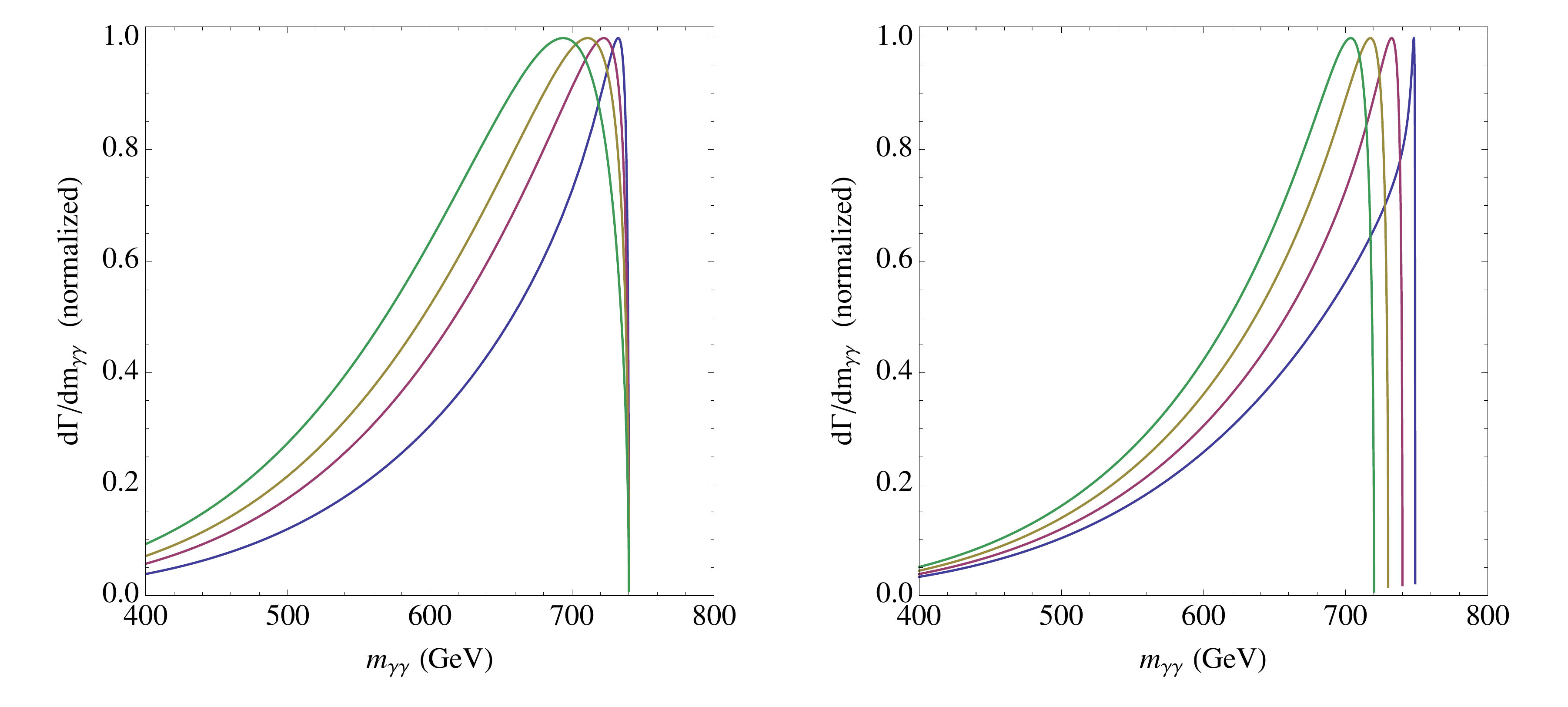}
\end{center}
\vspace*{-0.5cm}
\caption{Same as Fig.~\ref{fig:sc} but for fermion decay. The peaks are wider than for scalar decay, as explained in the text. }
\label{fig:fe} 
\end{figure*}

\subsection{Photon and Gluon Couplings}

As noted in \cite{falkowski}, $\phi$ decays are effectively mediated by the higher dimension operators
\eq{
 \frac{\alpha}{6\pi \Lambda}  \phi \, F_{\mu\nu} F^{\mu\nu} + \frac{\alpha_s}{6\pi \Lambda_s}  \phi \, G_{a\mu\nu} G^{a\mu\nu} \ ,
% + \frac{\alpha_s}{4\pi \Lambda_G}  \phi_1 G^a_{\mu\nu} G^{a\mu\nu}.
\label{eq:geniusoperator}
}
where in a weakly coupled theory $\Lambda$ and $\Lambda_s$ are of order the masses of heavy electrically charged and colored particles.  If, for example, these heavy particles are Dirac fermions then
\eq{
\frac{1}{\Lambda} = \sum_i \frac{k_i q_i^2}{m_i} \quad \textrm{and} \quad
\frac{1}{\Lambda_s} = \sum_i \frac{k_i T_i}{m_i} \ ,
}
where $i$ sums over heavy states.  Here $k_i$ denotes the Yukawa coupling of $\phi$ to a Dirac fermion of mass $m_i$, charge $q_i$, and color Dynkin index $T_i$.
 
 The decay rate of $\phi$ into diphoton is
\eq{
\Gamma(\phi \rightarrow \gamma\gamma) &=\frac{m_\phi^3}{4\pi}\left(\frac{\alpha }{6\pi \Lambda}\right)^2 \nn \\ &\simeq \textrm{6 keV} \times \left(\frac{m_\phi}{\textrm{750 GeV}} \right)^3 \left(\frac{\textrm{1 TeV}}{\Lambda} \right)^2 \ , \label{eq:gammagamma}
}
while the decay rate into digluon is
\eq{
\Gamma(\phi \rightarrow gg) &= 8\times \frac{m_\phi^3}{4\pi} \left(\frac{\alpha_s }{6\pi \Lambda_s}\right)^2 \nn \\ &\simeq \textrm{6 MeV} \times \left(\frac{m_\phi}{\textrm{750 GeV}} \right)^3 \left(\frac{\textrm{1 TeV}}{\Lambda_s} \right)^2 \ .\label{eq:gluongluon}
}
%XXX\draftnoteCC{Computed for $\alpha_s = 0.0913594$ and $\alpha^{-1} = 126.6$ from running couplings up to 750.}
The digluon rate is enhanced over the diphoton rate by a factor of $(\alpha_s/\alpha)^2$ together with a color multiplicity factor.
Notably, neither decay rate is remotely close to the $\sim$ 45 GeV width quoted by ATLAS, which is why we are interested in three-body decays in the first place.

With couplings to both photons and gluons, the relative branching ratio implied by \Eq{eq:gammagamma} and \Eq{eq:gluongluon} is
\eq{
\frac{\Gamma(\chi_1\to\chi_2 \gamma \gamma)}{\Gamma(\chi_1\to\chi_2 gg)} = \frac{\Gamma(\phi \rightarrow \gamma\gamma)}{\Gamma(\phi \rightarrow gg)} \simeq 10^{-3} \times \left( \frac{\Lambda_s}{\Lambda}\right)^2 \ .
\label{eq:Br}
}
For electrically charged and colored particles of similar mass scales, we see that the branching ratio between diphoton and digluon is of order one in a thousand, which is incredibly small.  To counterbalance this effect requires $\Lambda < \Lambda_s$, so the electrically charged particles are substantially lighter or more numerous than the colored ones.  This scenario is of course less constrained by LHC results since limits on colored production are far more stringent than for electroweak production.

In the subsequent sections we present simplified models characterizing the three-body decay into photons.  We split our discussion to the cases where the initial particle is a scalar or a fermion.

\subsection{Scalar Simplified Model}

%Concretely, if $\phi_1$ couples with 
%\eq{
%\frac{1}{\Lambda_F} &= \frac{2}{3}\sum_{\rm fermions} \frac{q^2}{M} + \frac{1}{3}\sum_{\rm bosons} \frac{q^2}{M}\\
%\frac{1}{\Lambda_G} &= \frac{2}{3}\sum_{\rm fermions} \frac{T}{M} + \frac{1}{3}\sum_{\rm bosons} \frac{T}{M},
%}

Let us define a simplified model for the three-body decay where $\chi_1$ and $\chi_2$ are scalars, interacting via
\eq{
%{\cal L} &=  
g \phi \chi_1 \chi_2 \ ,
%+ \frac{\alpha}{6\pi \Lambda}  \phi \, F_{\mu\nu} F^{\mu\nu}  \ ,
%+  \frac{\alpha_s}{6\pi \Lambda_s}  \phi \, G_{a\mu\nu} G^{a\mu\nu} \ ,
\label{eq:Lint}
}
where $g$ is a dimensionful cubic interaction. 
% This model is simply an augmentation by $\chi_1$ and $\chi_2$ of the simple $\phi$ effective theory described in many papers \cite{}. 
%While for generality we have included both the photon and gluon coupling, let us ignore the latter for now and consider processes involving just diphoton.
%The three-body 
%decay amplitude, summed over final state photon polarizations, is
%\eq{
%\sum_{\rm pols} |M|^2 &= 8g^2 \left( \frac{\alpha}{6\pi \Lambda} \right)^2 \frac{p^4}{(p^2 -m_\phi^2)^2} \ ,
%}
%where $p^2$ is the invariant mass of the diphoton system. 
%From this we compute the 
The three-body differential decay rate to photons is
\eq{
&\frac{d\Gamma({\chi_1\to\chi_2 \gamma\gamma})
}{dp^2} =\frac{g^2}{64\pi^3m_{\chi_1}^3} \left( \frac{\alpha}{6\pi \Lambda} \right)^2  \frac{p^4}{(p^2 -m_\phi^2)^2} \nn \\
& \quad \times  \sqrt{((m_{\chi_1} + m_{\chi_2})^2-p^2)((m_{\chi_1} - m_{\chi_2})^2-p^2)} \ ,
\label{eq:dGamma}
}
where we have defined
\eq{
p^2 &= m_{\gamma\gamma}^2 \ ,
}
to be the invariant mass of the diphoton system.
Note that as $p^2$ approaches $m_\phi^2$, the denominator approaches zero, yielding a singularity.  However, this pole is cut off by the zero in the numerator that arises when $p^2$ approaches $(m_{\chi_1} - m_{\chi_2})^2$.
This results in a peak for the diphoton spectrum that has a tunable width.

For illustrative purposes, in \Fig{fig:sc} we have plotted the diphoton spectrum for various choices of $m_\phi$, $m_{\chi_1}$, and $m_{\chi_2}$.  As shown, the observed width of $\sim$ 45 GeV is easily attainable. Moreover, the peaks are always asymmetric, with substantially more events to the left of the maximum than to the right.

Next, let us compute the total width into photons.
The expression in \Eq{eq:dGamma} is exact but unwieldy, so it will be convenient to consider the limit in which $m_{\chi_2} \ll m_{\chi_1}$, which as noted earlier is sufficient for the decay to yield highly energetic photons. In this limit the total width into photons is
\eq{
\Gamma({\chi_1\to\chi_2 \gamma\gamma}) &=\frac{g^2 m_{\phi}}{128 \pi^3 }  \left( \frac{\alpha}{6\pi \Lambda} \right)^2f(m_{\chi_1}/m_\phi) \ ,
}
where we have defined the function
\eq{
f(x) = \frac{x^2(x^2-6)+(4x^2-6)\log(1-x^2)}{x^3} \ .
}
For our kinematic regime, $f(m_{\chi_1} /m_\phi)$ is order one.  Thus, if the dominant decay channel is to photons, then the decay length is
\eq{
\tau \simeq 0.6 \textrm{~$\mu$m}  \times   \left(\frac{\textrm{750 GeV}}{m_\phi} \right)  \left(\frac{\textrm{100 GeV}}{g} \right)^2 \left(\frac{\Lambda}{\textrm{1 TeV}} \right)^2 \ ,
}
which is quite prompt on collider scales.

\subsection{Fermion Simplified Model} 

Next, consider a simplified model where $\chi_1$ and $\chi_2$ are Dirac fermions, interacting via
\eq{
%{\cal L} &= 
 y \phi (\bar\chi_1 \chi_2 +\bar\chi_2 \chi_1) \ ,
 %+ \frac{\alpha}{6\pi \Lambda}  \phi \, F_{\mu\nu} F^{\mu\nu}  \ , %+  \frac{\alpha_s}{6\pi \Lambda_s}  \phi \, G_{a\mu\nu} G^{a\mu\nu} \ ,
\label{eq:Lint}
}
where $y$ is a dimensionless Yukawa interaction. %The three-body decay amplitude, summed over final state photon and fermion polarizations and averaged over the initial fermion polarizations is
%\eq{
%\frac{1}{2} \sum_{\rm poles} |M|^2 &= 2y^2 \left( \frac{\alpha}{6\pi \Lambda} \right)^2 \frac{p^4((m_{\chi_1}+m_{\chi_2})^2-p^2)}{(p^2 -m_\phi^2)^2} \ .
%}
The three-body differential decay rate to photons is 
\eq{
&\frac{d\Gamma({\chi_1\to\chi_2 \gamma\gamma})}{dp^2} =\frac{y^2}{64\pi^3m_{\chi_1}^3} \left( \frac{\alpha}{6\pi \Lambda} \right)^2 \frac{p^4 ((m_{\chi_1}+m_{\chi_2})^2-p^2)}{(p^2 -m_\phi^2)^2}   \nn \\
&\quad  \times  \sqrt{((m_{\chi_1} + m_{\chi_2})^2-p^2)((m_{\chi_1} - m_{\chi_2})^2-p^2)} \ .
}
While this formula is similar to the one we obtained for the scalar model, there is a crucial difference: the numerator has a factor of $(m_{\chi_1}+m_{\chi_2})^2-p^2 $ which can cancel the singularity from the $\phi$ propagator.  For instance, in the limit that $m_{\chi_2}=0$ and $m_{\chi_1} = m_\phi$, the singularity is cancelled.  Hence, in this kinematic regime there is no resonant structure---merely an off-shell tail in the diphoton spectrum. 
Consequently, for a viable model we require that
\eq{
m_\phi &< m_{\chi_1}+ m_{\chi_2} \ ,
\label{eq:conditionF}
}
so the pole does not cancel. In view of Eq.~(\ref{eq:primarycondition}), this implies that $m_{\chi_2}\neq0$. This is an extra constraint that applies to the fermion model but not the scalar model.

In \Fig{fig:fe} we have plotted the diphoton spectrum from fermion three-body decays for various choices of $m_\phi$, $m_{\chi_1}$, and $m_{\chi_2}$.  As noted, for small $m_{\chi_2}$, the spectrum is more like an off-shell tail than a resonance.

Again restricting to $m_{\chi_2} \ll m_{\chi_1}$ for simplicity, we find that the total width into photons is
\eq{
\Gamma({\chi_1\to\chi_2 \gamma\gamma}) &=\frac{y^2m_\phi^6}{192 \pi^3 m_{\chi_1}^3 }  \left( \frac{\alpha}{6\pi \Lambda} \right)^2f(m_{\chi_1}/m_\phi) \ ,
}
where this time we define
\eq{
f(x) = 12x^2-12x^4+x^6+6(2-3x^2+x^4)\log(1-x^2) \ .
}
Since $f(m_{\chi_1} /m_\phi)$ is order one, the associated decay length is prompt,
\eq{
\tau \simeq 0.02 \textrm{ $\mu$m}\times y^{-2}\left(\frac{\textrm{750\,GeV}}{m_\phi} \right)^6 \left(\frac{m_{\chi_1}}{\textrm{750\,GeV}} \right)^3  \left(\frac{\Lambda}{\textrm{1\,TeV}} \right)^2, \label{eq:taufermion}
}
assuming here that the decay to photons is dominant.

\begin{figure}[t]
\vspace*{0.3cm}
\begin{center}
\includegraphics[width=.45\textwidth]{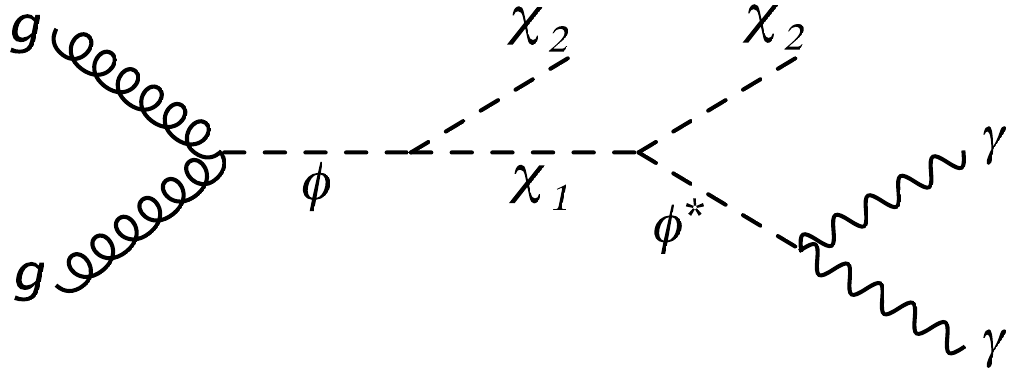}
\end{center}
\vspace*{-0.3cm}
\caption{Single production of a singlet scalar $\phi$ followed by cascade decay into dark matter and $\gamma\gamma$.}
\label{fig:single_production} 
\end{figure}

\section{Dark Matter Model}
\label{sec:DM}

We now present a simple weakly coupled model that explains the broad diphoton excess via three-body decays. As noted earlier, a diphoton spectrum arises only if $\phi$ is wider than $\chi_2$.  This is naturally accommodated if $\chi_2$ is stable dark matter, in which case $\chi_1$ and $\chi_2$ are odd under a dark matter stabilizing symmetry.  

The Lagrangian for this theory is simply
\eq{
{\cal L} &= g \phi \chi_1 \chi_2 +   \frac{\alpha}{6\pi \Lambda}  \phi \, F_{\mu\nu} F^{\mu\nu} + \frac{\alpha_s}{6\pi \Lambda_s}  \phi \, G_{a\mu\nu} G^{a\mu\nu}
\ ,
\label{eq:DML}
}
which is the usual effective theory for the diphoton excess~\cite{falkowski} augmented by $\chi_1$ and $\chi_2$.
Proton-proton collisions will produce cascade decays like the one depicted in \Fig{fig:single_production} to explain the diphoton excess.

\begin{figure*}
\begin{center}
\includegraphics[width=1\textwidth]{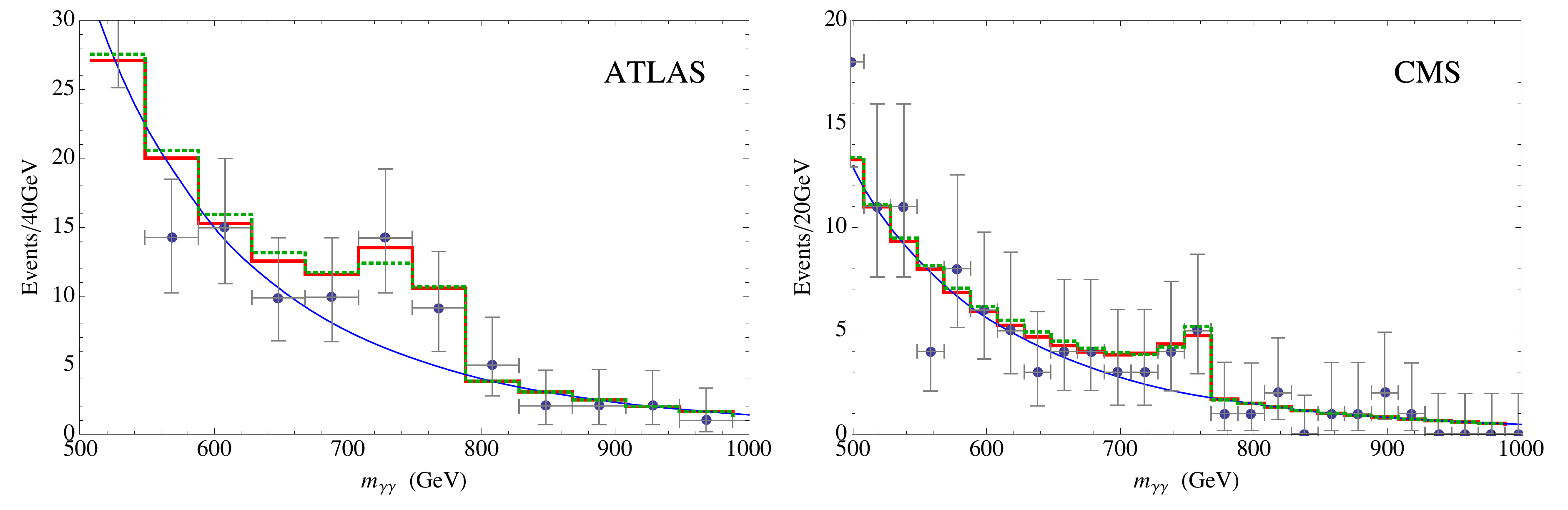}
\end{center}
\vspace*{-0.5cm}
\caption{A fit to the ATLAS and CMS diphoton excesses. 
The red histogram corresponds to the scalar model with $m_{\chi_1} = 769$ GeV, $m_\phi=780$ GeV, $m_{\chi_2}=0$ GeV.
The green histogram corresponds to the fermion model with $m_{\chi_1} = 764$ GeV, $m_\phi=760$ GeV, $m_{\chi_2}=5$ GeV.
The blue curves are the fits to the standard model background provided in the experimental papers. 
} 
\label{fig:fit_scalar}
\end{figure*} 

First, the $\phi$ particle is produced through its couplings to gluons with a cross-section
\eq{
\sigma(pp\rightarrow \phi ) &= \frac{\pi^2 \Gamma (\phi\to gg)}{8 m_\phi s} \int_{m_\phi^2/s}^{1} \frac{dx}{x} f_g(x) f_g\left( \frac{m_\phi^2/s}{x} \right) \ ,
\label{eq:gluonfusion}
}
where $f_g(x)$ is the gluon parton distribution function. 
%For concreteness, let us study a benchmark for which $m_\phi = 750$ GeV.  
%When the gluon coupling is sizable there are two competing effects: $\phi$ will be produced via gluon fusion, but at the same time $\phi$ will dominantly decay to gluons, decreasing its branching ratio to diphoton. While the $\phi$ width will increase, the state is still narrow relative to the observed diphoton excess.
For  $m_\phi = 750$ GeV, the production cross section is
\eq{
%\sigma(pp\rightarrow \phi ) &\sim 27 \textrm{ fb } \left( \frac{\textrm{1 TeV}}{\Lambda_s} \right)^2,
\sigma(pp\rightarrow \phi ) &\simeq 40 \textrm{ fb}\times  \left( \frac{\textrm{1 TeV}}{\Lambda_s} \right)^2,
}
which is quite sizable in spite of the fact that it is mediated by a higher dimension operator.

Once produced via gluon-gluon fusion, $\phi$ will decay two-body via $\phi \rightarrow \chi_1 \chi_2$, assuming $m_\phi > m_{\chi_1} + m_{\chi_2}$ so it is kinematically allowed.  Afterwards, the resulting $\chi_1$ has no recourse but to decay three-body back the way it came, as shown in \Fig{fig:single_production}. 
Taking into account the branching ratio dictated by \Eq{eq:Br}, we find 
\eq{
\sigma(pp\rightarrow  \gamma\gamma) &= \sigma(pp\rightarrow \phi) {\rm Br}(\phi\to \chi_1 \chi_2) {\rm Br}(\chi_1 \to \chi_2\gamma\gamma)   \nonumber \\
%&\quad\quad\quad 6  \textrm{ fb } \times \left[ 10^3 \left(\frac{\Lambda}{\textrm{1 TeV}}\right)^2 +  \left(\frac{\Lambda_s}{\textrm{1 TeV}}\right)^2 \right]^{-1} \ ,
& \!\!\!\!\!\!\!\!\!\!\!\!\!\!\!\!\!\!\!\!\!\!\!\!\!\!  \simeq 40  \textrm{ fb}  \times \left[ \left(\frac{\Lambda}{\textrm{30 GeV}}\right)^2 +  \left(\frac{\Lambda_s}{\textrm{1 TeV}}\right)^2 \right]^{-1} \ ,
}
for the final signal cross section.

The diphoton excess implies a signal cross-section between $\sim$ 5--10 fb.  To be concrete, a signal cross-section of $\sigma(pp\rightarrow  \gamma\gamma) \simeq $ 5 fb would require hierarchical values of $\Lambda \simeq$ 80 GeV and $\Lambda_s \simeq $ 1 TeV.  The higher dimension operators in \Eq{eq:DML} can thus be realized by coupling $\phi$ to vector-like quarks and leptons, with the latter substantially lighter. Note that since $\phi$ is not allowed to decay into these intermediate vector-like leptons, these states must be heavier than $\sim$ 375 GeV, which is half its mass. Therefore, to achieve small $\Lambda \lesssim 80$ GeV for order one Yukawa interaction, we need four to five copies of vector-like charged leptons to generate the signal.  Some of these multiplicity factors could arise if these leptons are multiplets of $SU(2)_L$.

In Fig.~\ref{fig:fit_scalar}, the red histogram is our fit to the excesses from ATLAS and CMS, choosing $m_{\chi_1} = 769$ GeV, $m_\phi=780$ GeV, and $m_{\chi_2} = 0$ GeV for the shape, and $\Lambda\simeq80$ GeV, $\Lambda_s=1$ TeV for the normalization.\footnote{Note that in the CMS analysis, the excess is shown with both of the photons reconstructed in the barrel region of the electromagnetic calorimeter (EBEB), which covers the region of $|\eta| < 1.479$ corresponding to $|\cos\theta| < 0.901$. Based on Monte Carlo simulation with MadGraph5~\cite{Alwall:2014hca}, this will reduce the signal by a factor of 0.63 and 0.71 for the scalar and fermion cases, respectively. }

Note that in every cascade decay, two $\chi_2$ dark matter particles are created.  Due to the somewhat compressed spectrum, these dark matter particles do not carry away an enormous amount of missing energy.  Nevertheless, this associated signal would be an interesting secondary prediction of this model.

\section{Heavy Quark and Lepton Model}

\label{sec:QL}

We now present an ultraviolet complete model for the diphoton excess which is weakly coupled and renormalizable.  This model is an extension of the standard model by a real singlet scalar $\phi$ plus a vector-like quark $Q$ and a vectorlike lepton $L$, which for simplicity we take to be $SU(2)_L$ singlets.  These states interact via
\eq{
{\cal L} &= y \phi (\bar q Q + \bar Q q) + k_Q \phi \bar Q Q + k_L \phi \bar L L\ .
}
We will consider when $q=b, t$ are the bottom and top quark, in which case $Q$ is effectively a heavy bottom or top partner.
As we will discuss shortly, the vector-like lepton $L$ must be electrically charged in order to produce a photon coupling to $\phi$.

The Yukawa couplings of $\phi$ to the $Q$ and $L$ radiatively generate gluon and photon couplings to $\phi$ with
\eq{
\frac{1}{\Lambda} =  3 q_Q^2\frac{k_Q}{m_Q} + q_L^2 \frac{k_{L}}{m_{L}} \quad \textrm{and} \quad \frac{1}{\Lambda_s} =\frac{k_Q}{2m_Q}  \ ,
\label{Bmodel}
}
in terms of the higher dimensional operators in \Eq{eq:geniusoperator}.

\begin{figure}[t]
\vspace*{0.3cm}
\begin{center}
\includegraphics[width=.37\textwidth]{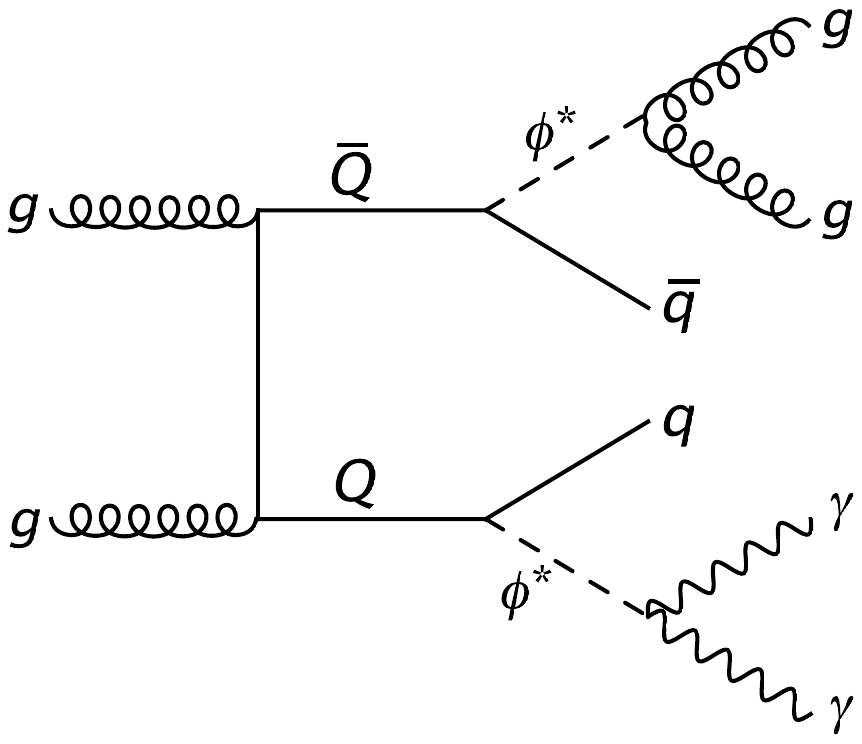}
\end{center}
\vspace*{-0.3cm}
\caption{Pair production of a vector-like quark $Q$ followed by cascade decays into $\gamma\gamma$ and $gg$, respectively, plus quarks.}
\label{fig:double_production} 
\end{figure}

In this model, $Q\bar Q$ are pair produced at the LHC and then each decays into a pair of photons or a pair of gluons plus a standard model quark $q$ at the parton level. The Feynman diagram is shown in Fig.~\ref{fig:double_production}. 
To fit the diphoton excesses observed by ATLAS and CMS, we require one of the $Q\bar Q$ pair to decay into photons and the other into gluons.  This is necessary since the diphoton signal arises from events with exactly two hard photons.
Given a target signal cross-section $\sigma(pp\rightarrow \gamma\gamma) $, we define the product of these branching ratios to be
\eq{
r& = {\rm Br} (Q\to q \gamma\gamma) {\rm Br} (Q\to  q g g) 
= \frac{\sigma(pp\rightarrow \gamma\gamma)}{2\sigma (pp\to Q\bar Q)}  \ .
\label{eq:rdef}
}
For $ {\rm Br} (Q\to q \gamma \gamma) \ll {\rm Br} ( Q\to q g g) $ this implies that
\eq{
r\simeq \frac{{\rm Br} (Q\to q \gamma\gamma)}{{\rm Br} ( Q\to q g g)} \ .
}
From Eqs.~(\ref{eq:Br}) and (\ref{Bmodel}), we then derive that
\begin{equation}\label{eq:vecBL}
q_L^2\frac{k_L}{m_L} \simeq \left( \sqrt{250r} - 3 q_Q^2 \right) \frac{k_Q}{m_Q} \ .
\end{equation}
%The current constraint on vector-like charged leptons from the 8 TeV LHC is $m_{Li} > 265$ GeV~\cite{Kumar:2015tna}. However, in order for $\phi$ to have a significant branching ratio to the diphoton channel, $m_{Li}$ must be above the $\phi$ decay threshold. Therefore, we have
%\begin{equation}
%m_{Li} > \frac{1}{2} m_\phi \approx 375{~\rm GeV} \ .
%\end{equation} 
%Therefore, one can see that Eq.~(\ref{eq:vecBL}) can be satisfied with one more vector-like quark with the coupling $k_L$ a little bit larger than $k_B$. 
Throughout, we assume a benchmark signal cross-section of $\sigma(pp\rightarrow \gamma\gamma) \simeq $ 5 fb to fit the diphoton excess.  
Note that the rate of heavy quark pair production is substantially higher at 13 TeV than 8 TeV, offering a reason why a diphoton excess was not observed in the first run of the LHC~\cite{Khachatryan:2015qba, Aad:2015kqa, Aad:2015mba}.
Next, let us discuss the case when $Q$ is a bottom partner or top partner, respectively.

\medskip

\noindent {\bf  Bottom Partner.} For the case $q=b$, the heavy quark $Q$ has the quantum numbers of the bottom.  In this scenario, we require $m_Q \sim m_{\gamma \gamma} + m_b \sim 750\,$GeV to fit the shapes of excesses from both ATLAS and CMS, as shown in the
green histograms in Fig.~\ref{fig:fit_scalar}. 

The pair production rate of 750 GeV vector-like quarks at the 13 TeV LHC is $\sigma (pp\to Q\bar Q)\simeq 260$ fb. This fixes an upper bound on the signal cross section, so from \Eq{eq:rdef}, $r\simeq 0.0096$.  To produce a high enough yield of diphoton, we require that $q_L^2 {k_L}/{m_L} \sim 1.2 {k_Q}/{m_Q}$ to be satisfied according to \Eq{eq:vecBL}, which is easily accommodated in a weakly coupled model.  For instance, we can take $k_Q$ to be small, thus decreasing the Yukawa coupling of $\phi$ to $Q$ and depleting the branching ratio of $\phi$ to gluons.  Since the production cross-section of $\phi$ is controlled by $Q$ pair production, this will not decrease the rate of $\phi$ production in proton-proton collisions.

In this scenario, the diphoton events in the signal are always accompanied by hard jets whose invariant mass will be close $\sim$ 750 GeV. This model also predicts a subpopulation of four-photon events with each pair reconstructing a $\sim$ 750 GeV bump, possibly observable at the LHC with higher luminosity. 

\medskip

\noindent {\bf  Top Partner.}  For the case $q=t$, the heavy quark $Q$ has the quantum numbers of the top.  Here we  need heavier $m_Q \sim m_{\gamma \gamma} + m_t \sim 920\,$GeV to fit the excesses so that an on-shell top can be produced. 
The pair production rate of 920 GeV vector-like quarks at the 13 TeV LHC is $\sigma (pp\to Q\bar Q)\simeq 70$ fb, implying that $r\simeq 0.036$.  
The observed diphoton rate requires $q_L^2 {k_L}/{m_L} \sim 1.7 {k_Q}/{m_Q}$, which is easily achievable in a weakly coupled theory.  

A potential issue here is that $Q$ can in principle decay via $Q\rightarrow (t^* \rightarrow W b) \phi$ through an off-shell top quark. If $W$ can be produced on-shell then the partial width for this channel is much greater than for the desired channel, $Q \rightarrow t (\phi^*\rightarrow \gamma\gamma)$. The way around this is to forbid this kinematically by requiring $m_Q -m_\phi <   m_W + m_b$ so that when $\phi$ is on-shell the residual energy is insufficient to produce an on-shell $W$ and $b$. In this case the leading decay of $Q$ via an off-shell $t$ is actually four-body via an off-shell $W$ and $t$, and thus highly suppressed.  The associated decay width is roughly
\begin{eqnarray}
\Gamma(Q\to  t^*\phi) &\simeq& \frac{y^2 g_2^4}{24576\pi^5} \frac{m_W^3}{m_t^2} \left(\frac{m_Q - m_\phi}{m_W}\right)^7 \nonumber \\
&\simeq& (0.02 \,\mu {\rm m})^{-1} \times  y^2  \left(\frac{m_Q - m_\phi}{50\,{\rm GeV}}\right)^7 \ ,
\end{eqnarray}
where $g_2\sim 0.65$ is the $SU(2)_L$ gauge coupling. One can see that $\Gamma(Q\to t^*\phi )$ is much smaller than the decay width of the channel with $t$ on-shell as shown in Eq.~(\ref{eq:taufermion}) if we choose $m_Q - m_\phi < 50$\,GeV. 

The diphoton signal produced in this case is similar to the bottom case discussed earlier. A key difference in the prediction is that here in addition to the two hard jets, one can also look for charged leptons and missing transverse energies associated with the diphoton.

%A potential issue here is that the decay to $b$ and $W$ through an off-shell top is much larger than that of $\phi$ given in Eqs.~(\ref{eq:gammagamma}) and (\ref{eq:gluongluon}), which violates the condition required by Eq.~(\ref{eq:conditionF}).
%This will make the decay $T \rightarrow t^* \phi$ via off-shell top quark and on-shell $\phi$ dominant and the diphoton resonance ends up too narrow. 

\section{Conclusions}

\label{sec:conclusions}

We have presented a weakly coupled mechanism that produces a broad, tunable resonance in the diphoton invariant mass spectrum.  This is implemented in theories where a three-body decay yields an off-shell state decaying to diphoton.  A generic prediction of these models is associated missing energy, jets, or leptons in events contributing to the diphoton excess.  

\bigskip

\noindent {\it Acknowledgements}: 
H.A. is supported by the Walter Burke Institute at Caltech and by DOE Grant No.~DE-SC0011632.
C.C. is supported by a DOE Early Career Award under Grant No.~DE-SC0010255 and a Sloan Research Fellowship.
Y.Z. is supported by the Gordon and Betty Moore Foundation through Grant \#776 to the Caltech Moore Center for Theoretical Cosmology and Physics, and by the DOE Grant Nos.~DE-FG02-92ER40701 and DE-SC0010255 .

\bibliographystyle{apsrev4-1}

\end{document}